\documentclass[aps,prb,reprint,superscriptaddress,floatfix]{revtex4-2}

\usepackage{amsmath,amssymb} 
\usepackage{graphicx} 
\usepackage{hyperref} 
\usepackage{xcolor} 
\usepackage{booktabs} 
\usepackage{siunitx} 
\usepackage{placeins} 

\newcommand{\PtNi}{\texorpdfstring{Pt$_{0.9}$Ni$_{0.1}$}{Pt0.9Ni0.1}} \newcommand{\degC}{\ensuremath{^{\circ}\mathrm{C}}}

\newcommand{\FigOneWidth}{0.43\textwidth} \newcommand{\FigTwoWidth}{0.95\textwidth} \newcommand{\FigThreeWidth}{0.5\textwidth}

\begin{document} \raggedbottom

\title{Substrate-dependent thermally driven morphological evolution of \PtNi{} thin films on sapphire and langasite}

\author{M.~Awais Fiaz} \email{MuhammadAwais.Fiaz@unh.edu} \affiliation{Department of Physics and Astronomy, University of New Hampshire, Durham, NH 03824, USA}

\author{M.~Greenslit} \affiliation{Department of Physics and Astronomy, University of Maine, Orono, ME 04469, USA}

\author{R.~J.~Lad} \affiliation{Department of Physics and Astronomy, University of Maine, Orono, ME 04469, USA}

\author{Mauricio~Pereira~da~Cunha} \affiliation{Department of Electrical and Computer Engineering, University of Maine, Orono, ME 04469, USA}

\author{Luke Doucette} \affiliation{Frontier Institute for Research in Sensor Technologies, University of Maine, Orono, ME 04469, USA}

\author{S.~M.~Hollen} \email{shawna.hollen@unh.edu} \affiliation{Department of Physics and Astronomy, University of New Hampshire, Durham, NH 03824, USA}

\date{August 10, 2026}

\hypersetup{hidelinks}

\begin{abstract} Pt-based thin films serve as high-temperature electrodes, sensing layers, and catalytic surfaces, where thermal coarsening can alter surface morphology and continuity. Alloying Pt with Ni can improve thermal stability, but the evolution of PtNi film morphology during annealing and the role of the substrate remain unclear. Here we study 100-nm-thick \PtNi{} films with a 10-nm Zr adhesion layer, deposited simultaneously on sapphire and langasite and annealed sequentially in vacuum from room temperature to 800\,\degC. Atomic force microscopy, feature segmentation, and height-distribution analysis show that films on both substrates remain dense and granular through 400\,\degC{} and restructure sharply between 400 and 600\,\degC~with feature area growing to $1.09\times10^{5}$~nm$^{2}$ on langasite and $1.67\times10^{5}$~nm$^{2}$ on sapphire. Above 600\,\degC{}, feature area continues to increase while mean height and RMS roughness decrease, revealing a change from combined lateral and vertical restructuring to predominantly lateral growth. At 800\,\degC{}, the surfaces are faceted and remain laterally connected, showing no signs of dewetting. By holding film composition, thickness, Zr adhesion layer, and thermal history fixed, we show that the substrate partly controls the extent of thermal restructuring and the final feature size. Substrate control and high temperature film connectivity are promising factors for high temperature applications. 
\end{abstract}

\maketitle

\section{Introduction} High temperature and extreme environment metallic alloys are critically important in high-temperature sensors, acoustic devices, and catalytic structures, but their function depends on retaining continuity, substrate adhesion, and a stable surface morphology during thermal exposure\cite{Frankel2015,Wang2008}. Heating activates atomic transport that can drive grain growth, agglomeration, and solid-state dewetting as the system reduces its surface and interfacial free energy\cite{Thompson2012,Galinski2010,Barmak2024,GonzalezGonzalez2011,Freund2004,Abadias2018}. 

Additionally, metal thin films on commonly used oxide substrates are metastable at high temperatures due to oxygen mobility\cite{Seifert2015}.  Two oxide substrates of interest for applications are sapphire, widely used as a support for thermally annealed metal films~\cite{Wen2022,Galinski2010,Sui2017}, and langasite, a high-temperature piezoelectric material used in acoustic-wave devices~\cite{DaCunha2023,DaCunha2008,Bardong2008,Aubert2011}.

In a supported thin-film geometry, the annealing response depends on the metal alloy and the interfacial reactions and stresses that develop during heating between the metal, adhesion layer, and substrate \cite{Moulzolf2014,Galinski2010,Abadias2018,Dai2018}. Pt-alloy thin films are a promising class,\cite{DaCunha2023,DaCunha2008,Frankel2015,Liu2017,Dai2018} and alloying Pt with Ni (PtNi) is shown to improve high-temperature film stability \cite{Frankel2015,Liu2017}. For pure Pt films, the oxide substrate was shown to influence grain-size evolution and surface morphology during annealing \cite{Wen2022,Galinski2010,Sui2017}, and the films dewetted at high temperatures, which is not suitable for many applications. For PtNi alloy thin films with a Zr adhesion layer (PtNi/Zr), the evolution of film morphology with annealing and the impact of the substrate choice is unknown.

Here we report morphological studies of annealed 100-nm-thick \PtNi{} films with a 10-nm Zr adhesion layer on sapphire (Al$_2$O$_3$) and langasite (La$_3$Ga$_5$SiO$_{14}$) substrates.  We deposited the films simultaneously and subjected both samples to the same stepwise annealing sequence from room temperature to 800\,\degC{}, keeping film composition, thickness, adhesion layer, and thermal history fixed on both substrates.  Three stages emerge from the thermal evolution, which we will call granular, restructuring, and faceted. Both films remain dense and granular through 400\,\degC{}. Between 400 and 600\,\degC{}, they undergo their strongest restructuring: the mean feature area increases 56-fold on langasite and 95-fold on sapphire, while mean height and RMS roughness reach their maxima. From 600 to 800\,\degC{}, projected feature area continues to increase while mean height and RMS roughness decrease from their maxima, marking a later stage of the morphological evolution. A stepwise Arrhenius-like treatment of the area increments supports this regime assignment, placing the strongest coarsening response in the 400--600\,\degC{} interval. At 800\,\degC{}, both films retain lateral connectivity and exhibit a faceted surface morphology, with sapphire reaching the larger final mean feature area. The connectivity of the films annealed to the highest temperatures is promising for applications, and the substrate dependence may give an opportunity for substrate engineering.

The H\"uttig and Tammann temperatures provide empirical reference scales for thermally activated atomic mobility~\cite{German1996,Aubert2011}. For a metal with melting temperature $T_{\rm m}$ in Kelvin, $T_{\rm H}\approx0.3T_{\rm m}$ marks the approximate onset of appreciable surface atomic mobility, whereas $T_{\rm T}\approx0.5T_{\rm m}$ corresponds to appreciable bulk atomic mobility. For Pt ($T_{\rm m}\approx2041$~K), $T_{\rm H}^{\rm Pt}\approx339\,\degC$ and $T_{\rm T}^{\rm Pt}\approx747\,\degC$; for Ni ($T_{\rm m}\approx1728$~K), $T_{\rm H}^{\rm Ni}\approx245\,\degC$ and $T_{\rm T}^{\rm Ni}\approx591\,\degC$. These values do not define sharp transition temperatures for a PtNi/Zr/substrate stack, but provide useful reference points for interpreting the annealing sequence. Metal--support interactions can further modify atomic mobility in supported systems~\cite{Dai2018,Galinski2010}.

The thermodynamic driving force for morphological restructuring is the reduction of the total free energy, which we represent schematically as \begin{equation} G = G_{\rm surf} + G_{\rm int} + G_{\rm gb} + G_{\rm stress}, \end{equation} where the terms represent surface, interface, grain-boundary, and stress contributions~\cite{Freund2004,Abadias2018,Thompson2012}. We use this energetic framework together with the H\"uttig and Tammann mobility scales to interpret the observed temperature regimes without assigning the sequential annealing response to a single microscopic transport mechanism.


\section{Methods}

We deposited 100-nm-thick \PtNi{} films simultaneously by e-beam co-evaporation at room temperature on optically polished langasite and c-cut sapphire substrates, following deposition of a 10-nm Zr adhesion layer. We annealed both films sequentially under high vacuum at a base pressure of approximately $2\times10^{-7}$~Torr at 200, 400, 600, and 800\,\degC{}, holding each temperature for 10~min. After each annealing step, we cooled the samples to room temperature in vacuum and characterized them ex~situ under ambient conditions.

We characterized surface morphology using a Bruker MultiMode atomic force microscope (AFM) operated in contact mode. AFM topographs were corrected by first-order plane subtraction in WSxM~\cite{Horcas2007} before quantitative analysis. We used scanning electron microscopy (SEM) to assess the large-area coverage and continuity of the as-deposited films and X-ray photoelectron spectroscopy (XPS) to characterize their surface composition and chemical state. The Pt~4$f$ and Ni~2$p$ core levels are consistent with metallic Pt and Ni, and the measured Pt:Ni atomic ratio is approximately 90:10~\cite{Moulder1995,NIST_XPS}.

We extracted feature statistics from the AFM topographs using ImageJ~\cite{Schneider2012}; representative contouring is shown in the Supplemental Material, Fig.~S2. The analysis provides projected area~$A$, equivalent circular diameter $d_{\rm eq}=2\sqrt{A/\pi}$, circularity $C=4\pi A/P^{2}$, and aspect ratio ${\rm AR}=L_{\rm major}/L_{\rm minor}$, where $P$ is the projected perimeter and $L_{\rm major}$ and $L_{\rm minor}$ are the major and minor axes. For RT--400\,\degC{}, hundreds of features are analyzed per image. At 600 and 800\,\degC{}, where the characteristic feature size is much larger, we analyzed multiple images and larger fields of view to improve sampling. Measurements at separated sample locations and with different AFM tips confirm the reproducibility of the observed trends and allow lateral connectivity to be assessed over multiple fields of view. Tip convolution can slightly broaden lateral dimensions and smooth sharp feature boundaries, but the systematic changes in projected area, circularity, and aspect ratio are reproduced across independent images and probes. The straight edges and polygonal outlines observed at 600 and 800\,\degC{} are consistent with a faceted surface morphology. Analysis of the AFM height distributions provides the mean height $\langle h\rangle$ and RMS roughness $R_q$. The same first-order plane subtraction was applied to all images before comparison of these vertical metrics. To compare the temperature dependence of the morphological evolution, we performed an effective Arrhenius-like analysis using the measured feature areas, as detailed in Appendix~\ref{app:derivation}. Each feature is represented by an equivalent-area circle, \begin{equation} A=\pi r^{2}, \qquad \langle r^{2}\rangle=\frac{\langle A\rangle}{\pi}. \end{equation}

For successive annealing steps, we calculated

\begin{equation} \Delta\langle r^{2}\rangle_i = \langle r^{2}\rangle_i- \langle r^{2}\rangle_{i-1}, \end{equation}

and assigned each increment to the higher annealing temperature~$T_i$. We plotted $\ln[\Delta\langle r^{2}\rangle]$, with $\Delta\langle r^{2}\rangle$ expressed in nm$^{2}$, against $1000/T_i$, where $T_i$ is in Kelvin. For adjacent points, the local slope is

\begin{equation} m_i= \frac{ \ln[\Delta\langle r^{2}\rangle_{i+1}] - \ln[\Delta\langle r^{2}\rangle_i] }{ (1000/T_{i+1})-(1000/T_i) }, \end{equation}

and the effective coarsening energy is defined as

\begin{equation} Q_{\rm eff}=-1000\,k_{\rm B}m_i, \end{equation}

where $k_{\rm B}$ is the Boltzmann constant. Because the measurements follow a cumulative sequence of 10-min anneals rather than independent isothermal time-series experiments, $Q_{\rm eff}$ describes the temperature sensitivity of the annealing intervals and is not interpreted as a microscopic activation energy.

\section{Results}

\begin{figure} \centering \includegraphics[width=\FigOneWidth]{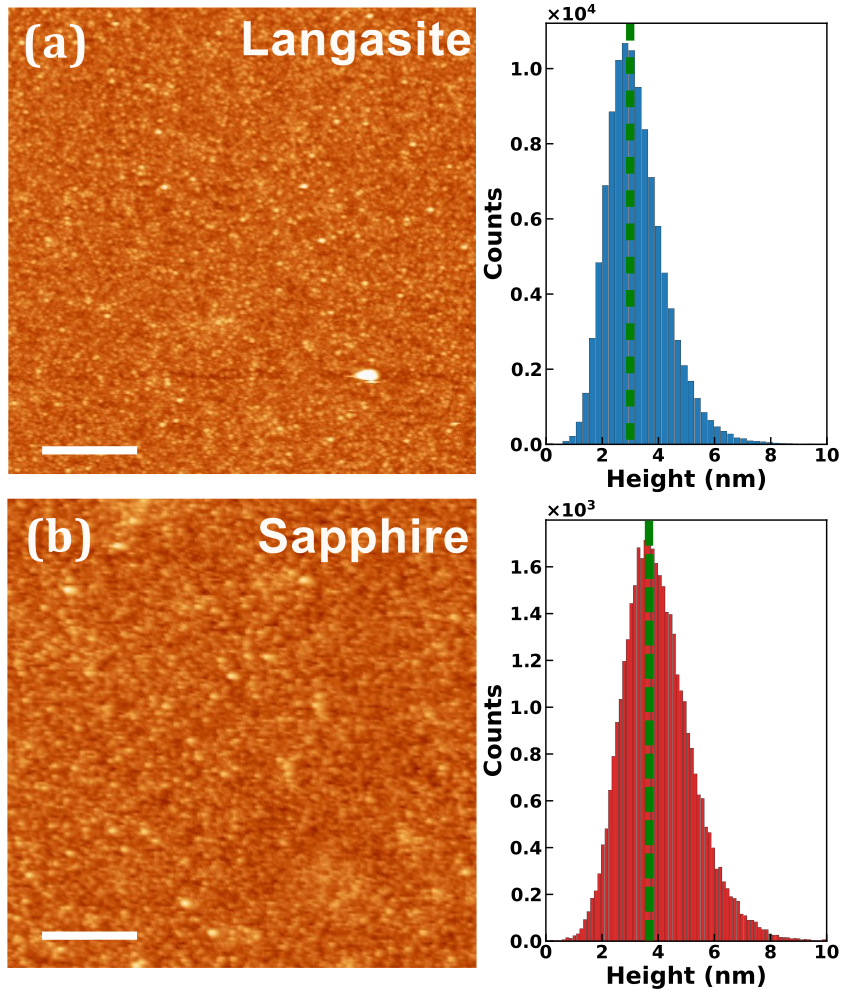} \caption{As-deposited \PtNi{}/Zr films on (a)~langasite and (b)~sapphire. AFM topographs and corresponding height distributions show continuous granular films before annealing. The dashed vertical lines indicate mean as-deposited heights of 3.02~nm for PtNi/Langasite and 5.52~nm for PtNi/Sapphire. The scale bars are 500~nm, and the vertical range is 15~nm for both topographs.} \label{fig:asdeposited} \end{figure}

\begin{table*} \centering \scriptsize \setlength{\tabcolsep}{4.8pt} \renewcommand{\arraystretch}{0.94} \caption{Morphology statistics of manually segmented Pt$_{0.9}$Ni$_{0.1}$ features and AFM height analysis. $T$ is the annealing temperature and $N$ is the number of segmented features. Projected area $\langle A\rangle$, equivalent circular diameter $d_{\rm eq}$, and circularity are obtained from segmented contours; mean height $\langle h\rangle$ and RMS roughness $R_q$ are obtained from the AFM height field. Uncertainties in $\langle A\rangle$ are standard errors of the mean.} \label{tab:data} \begin{ruledtabular} \begin{tabular}{llrccccc} Substrate & $T$ (\degC) & $N$ & $\langle A\rangle$~(nm$^2$) & $d_{\rm eq}$~(nm) & $\langle h\rangle$~(nm) & $R_q$~(nm) & Circ. \\ \hline PtNi/Langasite & RT  & 238 & $784 \pm 17$         & 31.6  & 3.02 & 1.05 & 0.761 \\                & 200 & 211 & $1053 \pm 22$        & 36.6  & 2.85 & 1.19 & 0.729 \\                & 400 & 711 & $1104 \pm 29$        & 37.5  & 3.35 & 1.12 & 0.815 \\                & 600 & 70  & $(61.70 \pm 2.0)\times10^{3}$  & 280.4 & 34.1 & 7.31 & 0.802 \\                & 800 & 28  & $(108.5 \pm 4.4)\times10^{3}$ & 371.7 & 5.66 & 2.57 & 0.744 \\[3pt] PtNi/Sapphire  & RT  & 392 & $332 \pm 7$          & 20.6  & 5.52 & 1.84 & 0.798 \\                & 200 & 316 & $335 \pm 12$         & 20.7  & 4.71 & 1.22 & 0.827 \\                & 400 & 261 & $1323 \pm 62$        & 41.0  & 3.16 & 1.13 & 0.738 \\                & 600 & 65  & $(126.0 \pm 3.7)\times10^{3}$ & 400.6 & 32.0 & 4.17 & 0.770 \\                & 800 & 36  & $(167.0 \pm 4.7)\times10^{3}$ & 461.1 & 5.80 & 2.23 & 0.787 \\ \end{tabular} \end{ruledtabular} \end{table*} 

The as-deposited films are continuous and densely granular on both substrates, but their initial morphologies differ (Fig.~\ref{fig:asdeposited}). PtNi/langasite has the larger mean projected feature area, $\langle A\rangle=784\pm17$~nm$^2$, compared with $332\pm7$~nm$^2$ for PtNi/sapphire. In contrast, sapphire has the greater mean height, 5.52~nm compared with 3.02~nm on langasite, and the greater RMS roughness, $R_q=1.84$~nm compared with 1.05~nm. XPS of the as-deposited films shows metallic Pt and Ni core levels and a Pt:Ni ratio close to 90:10 (Supplemental Material, Fig.~S1). A small number of crack-like surface features are also visible on both samples in Fig.~S1; their origin is not established by the present measurements.

\begin{figure*} \centering \includegraphics[width=\FigTwoWidth]{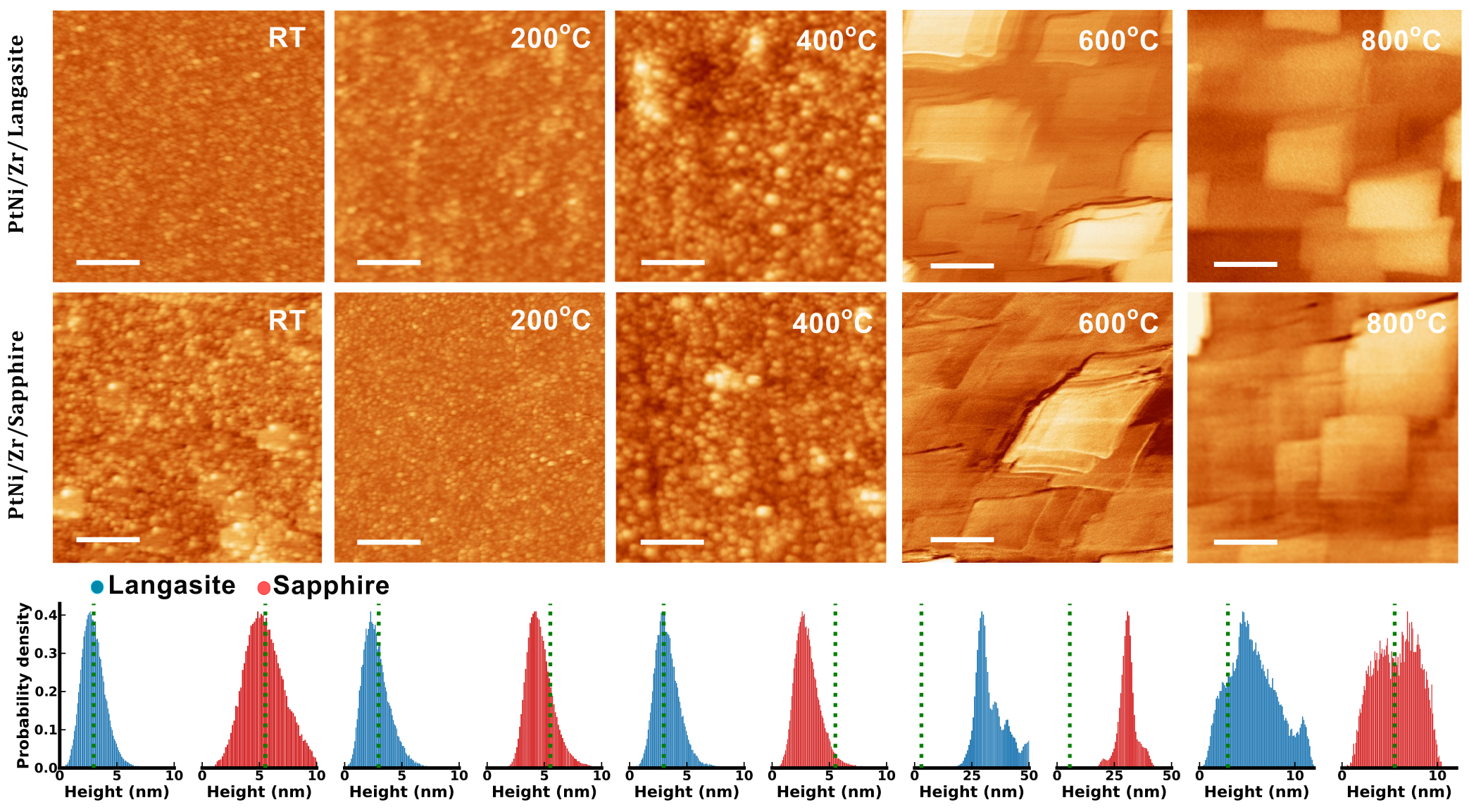} \caption{Room-temperature AFM topography of \PtNi{}/Zr films on langasite (top row) and sapphire (bottom row) acquired after successive vacuum annealing steps from RT to 800\,\degC{}. Both films remain dense and granular through the 400\,\degC{} anneal. The largest morphological change occurs between the 400 and 600\,\degC{} steps, with large straight-edged and polygonal features observed after annealing at 600 and 800\,\degC{}. Below each topograph is the corresponding surface-height distribution normalized to unit area. The vertical green dotted line marks the as-deposited mean height of each substrate and is held fixed across the corresponding row. Scale bars are 250~nm.} \label{fig:topography} \end{figure*}

Figure~\ref{fig:topography} separates the thermal evolution into three regimes. All AFM data were acquired at room temperature after the indicated annealing step. In Regime~I, from RT through the 400\,\degC{} anneal, both films retain a dense granular morphology, but their quantitative evolution differs. On langasite, the mean projected area increases modestly from $784\pm17$ to $1104\pm29$~nm$^2$, while the mean height and $R_q$ remain close to their initial values. On sapphire, the mean area is essentially unchanged after the 200\,\degC{} step and then increases to $1323\pm62$~nm$^2$ after the 400\,\degC{} step, approximately four times its as-deposited value. Over the same interval, the sapphire mean height decreases from 5.52 to 3.16~nm and $R_q$ decreases from 1.84 to 1.13~nm. The sapphire-supported film therefore exhibits lateral feature enlargement together with reduced vertical relief before the larger morphological change at higher temperature.

Regime~II spans the 400--600\,\degC{} annealing interval and contains the largest changes in the measured morphology. The mean projected area increases from $1104\pm29$ to $(61.7\pm2.0)\times10^{3}$~nm$^2$ on langasite and from $1323\pm62$ to $(126.0\pm3.7)\times10^{3}$~nm$^2$ on sapphire, corresponding to increases of approximately 56-fold and 95-fold, respectively. Mean height and roughness also reach their maxima after the 600\,\degC{} anneal: $\langle h\rangle=34.1$~nm and $R_q=7.31$~nm on langasite, compared with $\langle h\rangle=32.0$~nm and $R_q=4.17$~nm on sapphire. The corresponding AFM topographs show large straight-edged and polygonal features that are absent after the 400\,\degC{} anneal.

In Regime~III, between the 600 and 800\,\degC{} annealing steps, lateral feature growth continues while the vertical metrics decrease. The mean projected area reaches $(108.5\pm4.4)\times10^{3}$~nm$^2$ on langasite and $(167.0\pm4.7)\times10^{3}$~nm$^2$ on sapphire. Mean height decreases from 34.1 to 5.66~nm on langasite and from 32.0 to 5.80~nm on sapphire, while $R_q$ decreases from 7.31 to 2.57~nm and from 4.17 to 2.23~nm, respectively. The roughness after the 800\,\degC{} anneal is therefore well below its 600\,\degC{} maximum but remains above the corresponding 400\,\degC{} value on both substrates. Polygonal features consistent with a faceted surface morphology persist after the 800\,\degC{} anneal, and the surfaces remain laterally connected across the examined fields of view.

\begin{figure}
\centering \includegraphics[width=\FigThreeWidth]{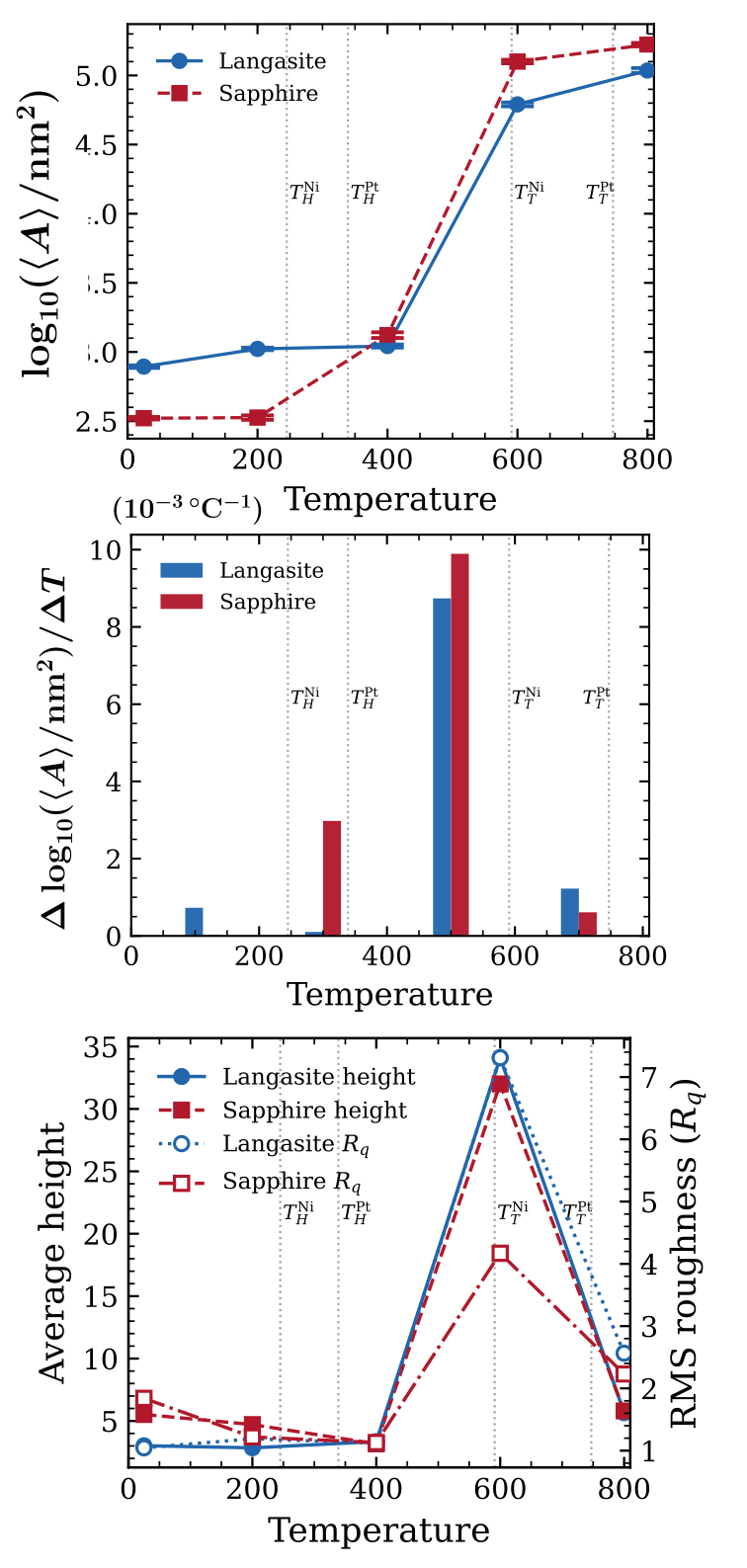} \caption{Temperature-dependent morphology metrics for PtNi/Langasite and PtNi/Sapphire. Top: $\log_{10}\langle A\rangle$ obtained from manually segmented contours, with the H\"uttig ($T_H$) and Tammann ($T_T$) reference temperatures for Ni and Pt shown as vertical markers. Middle: stepwise coarsening response, $\Delta\log_{10}(\langle A\rangle/\mathrm{nm}^{2})/\Delta T$, calculated between successive annealing temperatures. Bottom: mean height and RMS roughness obtained from AFM height distributions. The largest area-growth response and the maxima in height and roughness occur across the 400--600\,\degC{} interval. From 600 to 800\,\degC{}, projected area continues to increase while height and roughness decrease.} \label{fig:metrics} \end{figure}

Figure~\ref{fig:metrics} highlights the substrate dependence of these metrics. Sapphire reaches the larger mean projected feature area after both the 600 and 800\,\degC{} anneals, whereas langasite develops the larger roughness maximum after the 600\,\degC{} step.

\begin{figure} \centering \includegraphics[width=0.99\columnwidth]{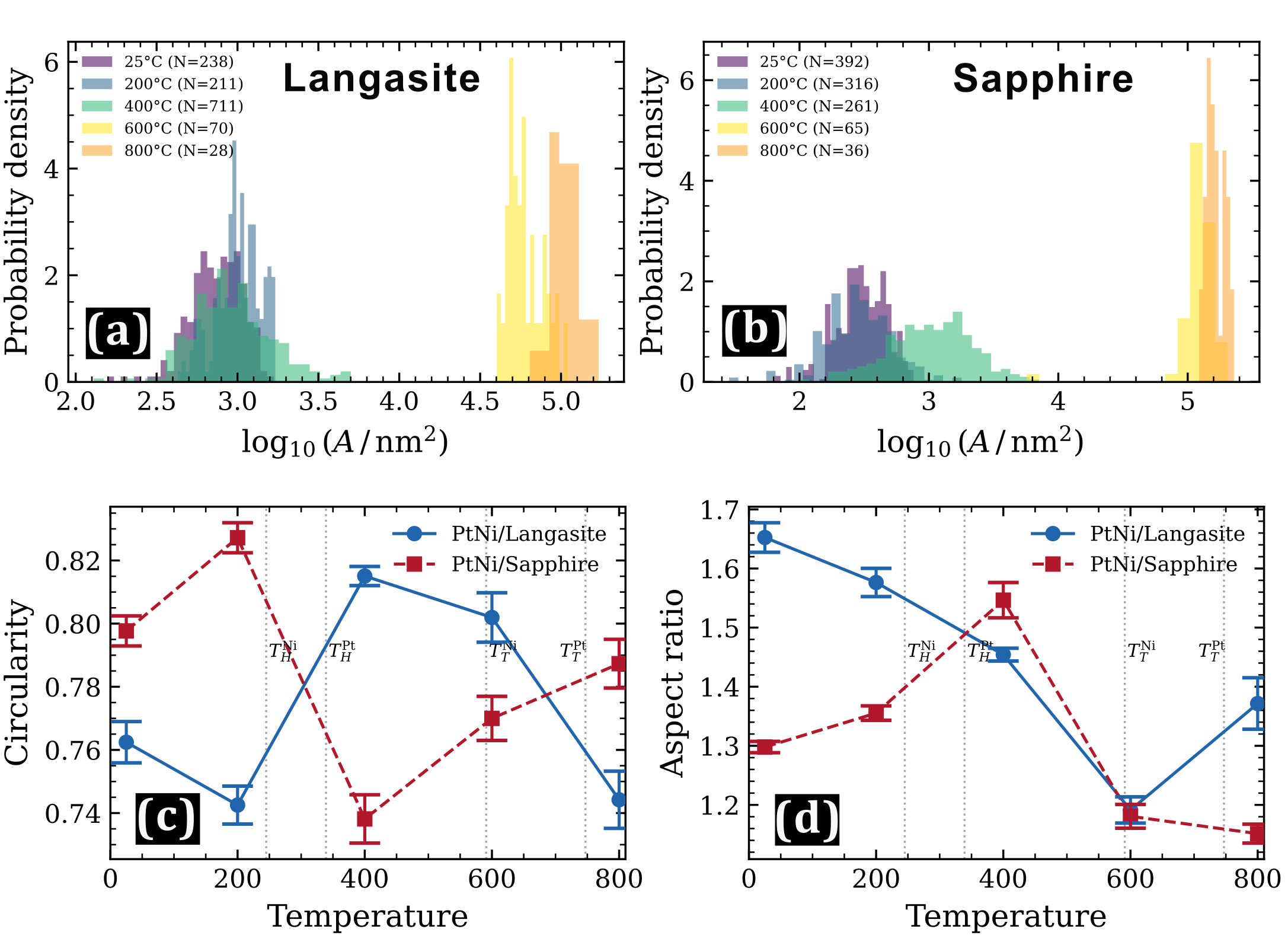} \caption{Probability-density histograms of segmented feature areas for a) PtNi/Langasite and b) PtNi/Sapphire. The distributions obtained after the 600\,\degC{} anneal are separated from the granular RT--400\,\degC{} distributions and extend into a large-feature range. c) Circularity and d) aspect ratio versus annealing temperature with H\"uttig and Tammann reference temperatures. Error bars represent the standard error of the mean. The non-monotonic shape metrics show that annealing changes feature geometry as well as projected area.} \label{fig:histograms} \end{figure}

The feature-area distributions in Figs.~\ref{fig:histograms}(a) and \ref{fig:histograms}(b), together with Supplemental Material Figs.~S4 and S5, show that the change in mean area reflects a population-level redistribution. Through the 400\,\degC{} anneal, the distributions remain concentrated in the small-feature range, with most projected areas near $10^{3}$~nm$^2$ or below, and the normalized distributions over this range are comparable in shape (Supplemental Material, Fig. S3). After the 600\,\degC{} anneal, the distributions are clearly separated from the low-temperature populations and extend into a large-feature range. Sapphire develops the broader high-area distribution after the 600\,\degC{} step and reaches the larger mean feature area after the 800\,\degC{} step. The feature-size distributions after the 600 and 800\,\degC{} annealing steps therefore differ between the two substrates despite the same annealing sequence.

Shape metrics provide complementary information about feature geometry. Circularity and aspect ratio vary non-monotonically with annealing temperature [Figs.~\ref{fig:histograms}(c) and \ref{fig:histograms}(d)]. The aspect ratio moves closer to unity after the 600\,\degC{} anneal on both substrates. After the 800\,\degC{} step, it increases again on langasite, whereas the sapphire-supported features remain closer to equiaxed. These changes show that annealing alters feature shape as well as characteristic size, which is consistent with the straight-edged polygonal morphology observed in the AFM topographs acquired after the 600 and 800\,\degC{} anneals.

\begin{figure}[!t] \centering \includegraphics[width=0.95\columnwidth]{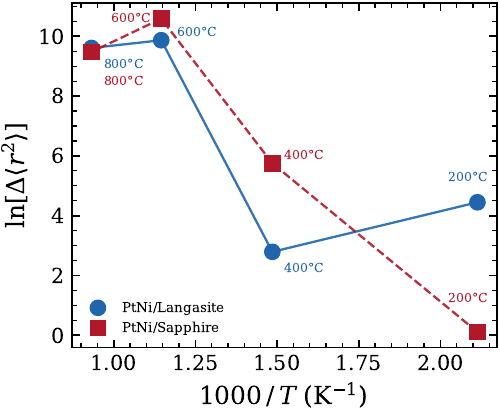} \caption{Effective Arrhenius-like analysis based on the stepwise increment $\Delta\langle r^{2}\rangle$, where $\langle r^{2}\rangle=\langle A\rangle/\pi$. Each increment is assigned to the higher annealing temperature, and $\ln[\Delta\langle r^{2}\rangle]$ is plotted against $1000/T$. The largest positive $Q_{\rm eff}$ response occurs across 400--600\,\degC{}. Above 600\,\degC{}, negative $Q_{\rm eff}$ values accompany continued area growth but a smaller stepwise increment. The analysis therefore distinguishes the main restructuring from the weaker high-temperature response.} \label{fig:arrhenius} \end{figure}

The effective Arrhenius-like analysis in Fig.~\ref{fig:arrhenius} provides a complementary measure of the stepwise temperature response. The largest positive local response occurs between the 400 and 600\,\degC{} assigned points, with $Q_{\rm eff}=1.79$~eV for PtNi/Langasite and 1.23~eV for PtNi/Sapphire. The low-temperature values are $-0.23$~eV on langasite and $0.77$~eV on sapphire, consistent with the comparatively limited area evolution before the principal restructuring. The 600--800\,\degC{} values are $-0.11$ and $-0.45$~eV, respectively. The negative high-temperature $Q_{\rm eff}$ values indicate a reduced incremental response after the principal restructuring even though the mean projected feature area continues to increase. Because the local slopes are obtained from successive area increments in the cumulative annealing sequence, the $Q_{\rm eff}$ values are used as regime descriptors rather than microscopic diffusion barriers.  \section{Discussion}

The \PtNi{}/Zr films on sapphire and langasite undergo their largest morphological change in the same 400--600\,\degC{} interval, but their lateral and vertical responses differ substantially. Through 400\,\degC{}, both films remain granular, with considerably stronger lateral enlargement on sapphire than on langasite. Between 400 and 600\,\degC{}, the response changes sharply: the mean projected feature area increases by approximately 56-fold on langasite and 95-fold on sapphire, while mean height and $R_q$ reach their maxima on both substrates. The simultaneous lateral growth and increase in surface roughness indicate extensive redistribution of material rather than a gradual enlargement of the original granular features. The separation of the 600\,\degC{} feature-area distributions from those measured through 400\,\degC{} supports the same interpretation.

The 400--600\,\degC{} interval lies above the H\"uttig temperatures of both Pt and Ni and reaches the Ni Tammann temperature (see Fig.~\ref{fig:metrics}), placing the observed transformation within a temperature range of substantially increased atomic mobility~\cite{German1996,Burke1952,Atkinson1988}. These empirical scales provide reference points for the thermal response of the constituent metals rather than transition temperatures for the \PtNi{}/Zr/substrate stack. The coincidence of the strong increase in feature area with the maxima in height and roughness shows that this interval involves both lateral coarsening and substantial redistribution of material normal to the surface.

Above 600\,\degC{}, the relation between lateral size and surface roughness reverses. Feature area continues to increase, while $R_q$ decreases from 7.31 to 2.57~nm on langasite and from 4.17 to 2.23~nm on sapphire between the 600 and 800\,\degC{} anneals. Continued coarsening therefore occurs predominantly through lateral redistribution rather than further increase in feature height. The straight-edged polygonal morphology observed after the 600 and 800\,\degC{} anneals is consistent with faceting and an increasing influence of anisotropic surface energy; similar faceted morphologies have been reported for thermally restructured Pt-based films~\cite{Vitos1998,Bonvicini2022}. The effective Arrhenius-like analysis gives the same distinction between temperature intervals: the largest positive $Q_{\rm eff}$ response occurs across 400--600\,\degC{}, followed by negative $Q_{\rm eff}$ values as the stepwise area increment decreases at higher temperature (Fig.~\ref{fig:arrhenius}).

The sapphire--langasite comparison shows that the choice of substrate changes the extent and form of restructuring and not the temperature interval in which the largest change occurs. At 600\,\degC{}, the mean projected feature area on sapphire is approximately twice that on langasite, while the roughness is lower (4.17 versus 7.31~nm). Greater lateral coarsening therefore does not correspond directly to greater surface roughness. Both supports are oxides, and the PtNi composition, film thickness, Zr adhesion layer, deposition conditions, and thermal history are the same; the different morphologies are therefore associated with the different PtNi/Zr/sapphire and PtNi/Zr/langasite interfaces. Differences in interfacial energy, adhesion, and thermally generated stress can modify the balance among surface, interface, grain-boundary, and stress energies and thereby alter the morphology reached during annealing~\cite{Galinski2010,Kovalenko2013,Rabkin2014,Abadias2018,Freund2004}. A similar dependence of annealed morphology on oxide substrate has been reported for pure Pt on sapphire and quartz by Wen~\emph{et al.}~\cite{Wen2022}, supporting the interpretation that the substrate modifies the extent of thermally driven restructuring. The present results establish this substrate dependence for an alloyed PtNi/Zr stack during sequential high-vacuum annealing.

Despite substantial coarsening and the development of faceted features, both films remain laterally connected across the examined AFM regions after the 800\,\degC{} anneal. Pt/Ta metallization on langasite has been reported to degrade near 900\,\degC{}~\cite{Aubert2011}, although the material system and thermal exposure differ from the present 10-min vacuum anneals to 800\,\degC{}. Based on the high-temperature behavior reported for PtNi-containing films, we expect the present \PtNi{}/Zr system to provide greater morphological stability than Pt/Ta. Connectivity and morphological stability describe different aspects of the thermal response: a film can remain connected while its characteristic feature size, roughness, and surface geometry change substantially. The retained connectivity is consistent with the improved high-temperature stability reported for PtNi-based metallization~\cite{Liu2017,Frankel2015}, while the Zr adhesion layer can further influence the thermal response of Pt-based films~\cite{Moulzolf2014}. Most importantly, the different morphologies reached on sapphire and langasite show that the high-temperature response is a property of the complete PtNi/Zr/oxide architecture, not of the PtNi alloy alone. \section{Conclusions}

We quantified the thermal evolution of 100-nm-thick \PtNi{} films with a 10-nm Zr adhesion layer on sapphire and langasite during sequential high-vacuum annealing up to 800\,\degC{}. Both films remain dense and granular through 400\,\degC{} and undergo their strongest morphological change between 400 and 600\,\degC{}, where projected feature area increases sharply and mean height and RMS roughness reach their maxima. The feature-area distributions measured after the 600\,\degC{} anneal represent much larger feature sizes, and the stepwise Arrhenius-like analysis places the strongest coarsening response in the same interval. From 600 to 800\,\degC{}, projected feature area continues to increase while height and roughness decrease. After the 800\,\degC{} anneal, both films exhibit a faceted surface morphology and remain laterally connected across the examined regions.

Under the same film composition, thickness, Zr adhesion layer, and thermal history, sapphire reaches a larger final mean feature area ($1.67\times10^{5}$~nm$^{2}$) compared with langasite ($1.09\times10^{5}$~nm$^{2}$). The different morphologies reached on the two oxide substrates show that the substrate partly controls the extent of thermal restructuring and the characteristic feature size during annealing. These results show the promise of PtNi/Zr stacks on sapphire and langasite substrates for high-temperature electrodes, sensors, and catalytic surfaces. Moreover, they show that in addition to alloy composition and adhesion-layer structure, the substrate selection is an important design variable in high temperature applications. 

\section*{Appendix: Derivation of the effective area-based coarsening relation} \label{app:derivation}

\noindent As a phenomenological reference, normal grain-growth kinetics are commonly expressed as~\cite{Burke1952,Atkinson1988,German1996} \begin{equation} D^{n}-D_{0}^{n}=K(T)t, \label{eq:grain_growth} \end{equation} where $D$ is a characteristic lateral length scale, $n$ is a growth exponent, $K(T)$ is a temperature-dependent rate coefficient, and $t$ is the annealing time. We use this relation only to construct an area-based measure of the observed coarsening; the segmented AFM features are not assumed to represent individual crystallographic grains.

For each segmented feature, we define an equivalent-area radius $r$ through \begin{equation} A=\pi r^{2}, \qquad \langle r^{2}\rangle=\frac{\langle A\rangle}{\pi}. \label{eq:equiv_radius} \end{equation} At each annealing condition, the characteristic lateral length in Eq.~\eqref{eq:grain_growth} is represented by \begin{equation} D=\alpha\sqrt{\langle r^{2}\rangle}, \label{eq:D_r_relation} \end{equation} where $\alpha$ is a temperature-independent geometrical factor set by the convention used to define $D$. For example, if $D$ is defined as the equivalent circular diameter associated with the mean projected area, $\alpha=2$. Thus, the area-derived radius is not assumed to be identical to a crystallographic grain size; it provides a consistent characteristic lateral length scale proportional to $D$.

We take $n=2$, corresponding to the classical parabolic growth law~\cite{Burke1952,Atkinson1988,German1996}. For two successive annealing conditions, Eq.~\eqref{eq:grain_growth} then gives \begin{equation} D_i^{2}-D_{i-1}^{2} = \alpha^{2} \left( \langle r^{2}\rangle_i-\langle r^{2}\rangle_{i-1} \right) = K(T_i)t. \label{eq:D_r_increment} \end{equation} Defining \begin{equation} \Delta\langle r^{2}\rangle_i = \langle r^{2}\rangle_i-\langle r^{2}\rangle_{i-1} = \frac{ \langle A\rangle_i-\langle A\rangle_{i-1} }{\pi}, \label{eq:delta_r2} \end{equation} we obtain \begin{equation} \Delta\langle r^{2}\rangle_i = K_r(T_i)t, \qquad K_r(T)=\frac{K(T)}{\alpha^{2}}. \label{eq:Kr} \end{equation} The geometrical factor $\alpha^{-2}$ therefore changes only the prefactor and does not affect the temperature dependence or the slope of the logarithmic relation below.

For an Arrhenius temperature dependence of the rate coefficient~\cite{German1996}, \begin{equation} K_r(T)=K_{r,0} \exp\left(-\frac{Q}{k_{\rm B}T}\right), \label{eq:arrhenius_K} \end{equation} a set of independent isothermal anneals of equal duration would satisfy \begin{equation} \ln\left( \frac{\Delta\langle r^{2}\rangle}{r_{\rm ref}^{2}} \right) = C-\frac{Q}{k_{\rm B}T}, \label{eq:arrhenius_r2} \end{equation} where $r_{\rm ref}^{2}=1$~nm$^{2}$ is a fixed reference area and \begin{equation} C= \ln\left( \frac{K_{r,0}t}{r_{\rm ref}^{2}} \right) \end{equation} contains the temperature-independent factors. This normalization is equivalent to using the numerical value of $\Delta\langle r^{2}\rangle$ expressed in nm$^{2}$ and therefore does not affect the slope.

Introducing \begin{equation} x=\frac{1000}{T}, \label{eq:x_1000T} \end{equation} as used in Fig.~\ref{fig:arrhenius}, Eq.~\eqref{eq:arrhenius_r2} becomes \begin{equation} \ln\left( \frac{\Delta\langle r^{2}\rangle}{r_{\rm ref}^{2}} \right) = C-\frac{Q}{1000\,k_{\rm B}}x. \label{eq:arrhenius_x} \end{equation} For two adjacent points, the local slope with respect to $1000/T$ is \begin{equation} m_i= \frac{ \ln[\Delta\langle r^{2}\rangle_{i+1}] - \ln[\Delta\langle r^{2}\rangle_i] }{ (1000/T_{i+1})-(1000/T_i) }, \label{eq:local_slope} \end{equation} which gives \begin{equation} Q_{\rm eff}=-1000\,k_{\rm B}m_i. \label{eq:Qeff} \end{equation}

The choice $n=2$ provides a direct mapping between the classical parabolic growth relation and the experimentally measured projected-area increments. The present sequential annealing data do not provide an independent isothermal time series from which a growth exponent can be determined, so $n=2$ is used as a reference exponent rather than as an assignment of a microscopic coarsening mechanism. Other coarsening mechanisms can exhibit different growth laws~\cite{Lifshitz1961,Wagner1961}. In addition, the experimental points follow a cumulative sequence of 10-min anneals rather than independent isothermal measurements. The resulting $Q_{\rm eff}$ values therefore quantify the relative temperature response of successive annealing intervals rather than activation energies for a specific atomic transport process. 

\section*{Data Availability}
\noindent The data that support the findings of this study are available from the corresponding authors upon reasonable request. 

\begin{acknowledgments} \noindent This material is based upon work supported by the U.S. Department of Energy, Office of Science, Office of Basic Energy Sciences Established Program to Stimulate Competitive Research (EPSCoR) under Award Number DE-SC0021981. \end{acknowledgments} 

\noindent \textbf{Disclaimer:} This report was prepared as an account of work sponsored by an agency of the United States Government. Neither the United States Government nor any agency thereof, nor any of their employees, makes any warranty, express or implied, or assumes any legal liability or responsibility for the accuracy, completeness, or usefulness of any information, apparatus, product, or process disclosed, or represents that its use would not infringe privately owned rights. Reference herein to any specific commercial product, process, or service by trade name, trademark, manufacturer, or otherwise does not necessarily constitute or imply its endorsement, recommendation, or favoring by the United States Government or any agency thereof. The views and opinions of authors expressed herein do not necessarily state or reflect those of the United States Government or any agency thereof. 

\noindent\textbf{Declaration of generative AI and AI-assisted technologies in the manuscript preparation process:}
During the preparation of the first draft of this work, the first author used chatGPT for grammar and English language assistance. The authors reviewed and edited the output as needed and take full responsibility for the content of the published article. 

\bibliography{references}

@inproceedings{DaCunha2023,
  author    = {{Mauricio Pereira da Cunha}},
  title     = {High-temperature harsh-environment {SAW} sensor technology},
  booktitle = {2023 IEEE International Ultrasonics Symposium (IUS)},
  pages     = {1--10},
  year      = {2023},
  doi       = {10.1109/IUS51837.2023.10306640}
}

@inproceedings{DaCunha2008,
  author    = {{Mauricio Pereira da Cunha} and Moonlight, T. and Lad, R. and Frankel, D. and Bernhard, G.},
  title     = {High temperature sensing technology for applications up to 1000\,{$^\circ$C}},
  booktitle = {2008 IEEE Sensors},
  pages     = {752--755},
  year      = {2008},
  doi       = {10.1109/ICSENS.2008.4716550}
}

@article{Frankel2015,
  author  = {Frankel, D. J. and Moulzolf, S. C. and {Mauricio Pereira da Cunha} and Lad, R. J.},
  title   = {Influence of composition and multilayer architecture on electrical conductivity of high temperature {Pt}-alloy films},
  journal = {Surf. Coat. Technol.},
  volume  = {284},
  pages   = {215--221},
  year    = {2015},
  doi     = {10.1016/j.surfcoat.2015.08.074}
}

@article{Moulzolf2014,
  author  = {Moulzolf, S. C. and Frankel, D. J. and {Mauricio Pereira da Cunha} and Lad, R. J.},
  title   = {High temperature stability of electrically conductive {Pt--Rh/ZrO$_2$} and {Pt--Rh/HfO$_2$} nanocomposite thin film electrodes},
  journal = {Microsyst. Technol.},
  volume  = {20},
  pages   = {523--531},
  year    = {2014},
  doi     = {10.1007/s00542-013-1974-x}
}

@article{Aubert2011,
  author={Aubert, T. and Elmazria, O. and Assouar, B. and Bouvot, L. and Hehn, M. and Weber, S. and Oudich, M. and Gen{\`e}ve, D.},
  title={Behavior of platinum/tantalum as interdigital transducers for {SAW} devices in high-temperature environments},
  journal={IEEE Trans. Ultrason. Ferroelectr. Freq. Control},volume={58},pages={603--610},year={2011},
  doi={10.1109/TUFFC.2011.1843}}

@inproceedings{Bardong2008,
  author={Bardong, J. and Schulz, M. and Schmitt, M. and Shrena, I. and Eisele, D. and Mayer, E. and Reindl, L. M. and Fritze, H.},
  title={Precise measurements of {BAW} and {SAW} properties of langasite in the temperature range from 25\,{$^\circ$C} to 1000\,{$^\circ$C}},
  booktitle={2008 IEEE International Frequency Control Symposium},pages={326--331},year={2008},
  doi={10.1109/FREQ.2008.4623013}}

@article{Seifert2015,
  author={Seifert, M. and Rane, G. K. and Kirbus, B. and Menzel, S. B. and Gemming, T.},
  title={Surface effects and challenges for application of piezoelectric langasite substrates in {SAW} devices caused by high temperature annealing under high vacuum},
  journal={Materials},volume={8},pages={8868--8876},year={2015},
  doi={10.3390/ma8125497}}

@article{Bonvicini2022,
  author  = {Bonvicini, Stephanie Nicole and Fu, Bo and
             Fulton, Alison Joy and Jia, Zhitai and Shi, Yujun},
  title   = {Formation of {Au}, {Pt}, and bimetallic {Au--Pt}
             nanostructures from thermal dewetting of single-layer
             or bilayer thin films},
  journal = {Nanotechnology},
  volume  = {33},
  number  = {23},
  pages   = {235604},
  year    = {2022},
  doi     = {10.1088/1361-6528/ac5a83}
}

@article{Dai2018,
  author  = {Dai, Yunqian and Lu, Ping and Cao, Zhenming and
             Campbell, Charles T. and Xia, Younan},
  title   = {The physical chemistry and materials science behind
             sinter-resistant catalysts},
  journal = {Chemical Society Reviews},
  volume  = {47},
  number  = {12},
  pages   = {4314--4331},
  year    = {2018},
  doi     = {10.1039/C7CS00650K}
}

@article{Abadias2018,
  author={Abadias, G. and Chason, E. and Keckes, J. and Sebastiani, M. and Thompson, G. B. and Barthel, E. and Doll, G. L. and Murray, C. E. and Stoessel, C. H. and Martinu, L.},
  title={Stress in thin films and coatings: Current status, challenges, and prospects},
  journal={J. Vac. Sci. Technol. A},volume={36},pages={020801},year={2018},
  doi={10.1116/1.5011790}}

@book{Freund2004,
  author={Freund, L. B. and Suresh, S.},
  title={Thin Film Materials: Stress, Defect Formation and Surface Evolution},
  publisher={Cambridge University Press},year={2004},
  doi={10.1017/CBO9780511754715}}

@article{Thompson2012,
  author={Thompson, C. V.},
  title={Solid-state dewetting of thin films},
  journal={Annu. Rev. Mater. Res.},volume={42},pages={399--434},year={2012},
  doi={10.1146/annurev-matsci-070511-155048}}

@article{Galinski2010,
  author={Galinski, H. and Ryll, T. and Elser, P. and Rupp, J. L. M. and Bieberle-H{\"u}tter, A. and Gauckler, L. J.},
  title={Agglomeration of {Pt} thin films on dielectric substrates},
  journal={Phys. Rev. B},volume={82},pages={235415},year={2010},
  doi={10.1103/PhysRevB.82.235415}}

@article{Sui2017,
  author={Sui, M. and Li, M.-Y. and Kunwar, S. and Pandey, P. and Zhang, Q. and Lee, J.},
  title={Effects of annealing temperature and duration on the morphological and optical evolution of self-assembled {Pt} nanostructures on c-plane sapphire},
  journal={PLoS ONE},volume={12},pages={e0177048},year={2017},
  doi={10.1371/journal.pone.0177048}}

@article{Kovalenko2013,
  author={Kovalenko, O. and Greer, J. R. and Rabkin, E.},
  title={Solid-state dewetting of thin iron films on sapphire substrates controlled by grain boundary diffusion},
  journal={Acta Mater.},volume={61},pages={3148--3156},year={2013},
  doi={10.1016/j.actamat.2013.01.062}}

@article{Rabkin2014,
  author={Rabkin, E. and Amram, D. and Alster, E.},
  title={Solid state dewetting and stress relaxation in a thin single crystalline {Ni} film on sapphire},
  journal={Acta Mater.},volume={74},pages={30--38},year={2014},
  doi={10.1016/j.actamat.2014.04.020}}

@article{Wen2022,
  author  = {J. Wen and J. Li and J. He and Y. Chen and X. Yan and Q. Guo
             and Q. Zhou and L. Wei and J. Sun and H. Guo},
  title   = {Morphological evolution of {Pt}-films on sapphire and quartz
             substrates at various temperatures: An experimental and
             molecular dynamics study},
  journal = {Appl. Surf. Sci.},
  volume  = {588},
  pages   = {152937},
  year    = {2022},
  doi     = {10.1016/j.apsusc.2022.152937}
}

@article{Liu2017,
  author={Liu, K.-Y. and Yoon, Y. J. and Lee, S. H. and Su, P.-C.},
  title={Sputtered nanoporous {PtNi} thin film cathodes with improved thermal stability for low temperature solid oxide fuel cells},
  journal={Electrochim. Acta},volume={247},pages={558--563},year={2017},
  doi={10.1016/j.electacta.2017.07.064}}

@article{Wang2008,
  author={Wang, X. and Huang, H. and Holme, T. and Tian, X. and Prinz, F. B.},
  title={Thermal stabilities of nanoporous metallic electrodes at elevated temperatures},
  journal={J. Power Sources},volume={175},pages={75--81},year={2008},
  doi={10.1016/j.jpowsour.2007.09.066}}

@article{Vitos1998,
  author={Vitos, L. and Ruban, A. V. and Skriver, H. L. and Koll{\'a}r, J.},
  title={The surface energy of metals},
  journal={Surf. Sci.},volume={411},pages={186--202},year={1998},
  doi={10.1016/S0039-6028(98)00363-X}}

@article{Burke1952,
  author={Burke, J. E. and Turnbull, D.},
  title={Recrystallization and grain growth},
  journal={Prog. Met. Phys.},volume={3},pages={220--292},year={1952},
  doi={10.1016/0502-8205(52)90009-9}}

@article{Atkinson1988,
  author={Atkinson, H. V.},
  title={Theories of normal grain growth in pure single phase systems},
  journal={Acta Metall.},volume={36},pages={469--491},year={1988},
  doi={10.1016/0001-6160(88)90079-X}}

@book{German1996,
  author={German, R. L.},
  title={Sintering Theory and Practice},
  publisher={Wiley},address={New York},year={1996}}

@article{Lifshitz1961,
  author={Lifshitz, I. M. and Slyozov, V. V.},
  title={The kinetics of precipitation from supersaturated solid solutions},
  journal={J. Phys. Chem. Solids},volume={19},pages={35--50},year={1961},
  doi={10.1016/0022-3697(61)90054-3}}

@article{Wagner1961,
  author={Wagner, C.},
  title={{Theorie der Alterung von Niederschl{\"a}gen durch Uml{\"o}sen}},
  journal={Z. Elektrochem.},volume={65},pages={581--591},year={1961}}

@article{Barmak2024,
  author={Barmak, K. and Rickman, J. M. and Patrick, M. J.},
  title={Advances in experimental studies of grain growth in thin films},
  journal={JOM},volume={76},pages={3622--3636},year={2024},
  doi={10.1007/s11837-024-06475-9}}

@article{GonzalezGonzalez2011,
  author={Gonz{\'a}lez-Gonz{\'a}lez, A. and Alonzo-Medina, G. M. and Oliva, A. I. and Polop, C. and Saced{\'o}n, J. L. and Vasco, E.},
  title={Morphology evolution of thermally annealed polycrystalline thin films},
  journal={Phys. Rev. B},volume={84},pages={155450},year={2011},
  doi={10.1103/PhysRevB.84.155450}}

@article{Horcas2007,
  author={Horcas, I. and Fern{\'a}ndez, R. and G{\'o}mez-Rodr{\'i}guez, J. M. and Colchero, J. and G{\'o}mez-Herrero, J. and Baro, A. M.},
  title={{WSXM}: A software for scanning probe microscopy and a tool for nanotechnology},
  journal={Rev. Sci. Instrum.},volume={78},pages={013705},year={2007},
  doi={10.1063/1.2432410}}

@article{Schneider2012,
  author={Schneider, C. A. and Rasband, W. S. and Eliceiri, K. W.},
  title={{NIH Image} to {ImageJ}: 25 years of image analysis},
  journal={Nat. Methods},volume={9},pages={671--675},year={2012},
  doi={10.1038/nmeth.2089}}

@book{Moulder1995,
  author={Moulder, J. F. and Stickle, W. F. and Sobol, P. E. and Bomben, K. D.},
  title={Handbook of {X}-ray Photoelectron Spectroscopy},
  publisher={Physical Electronics},address={Eden Prairie, MN},year={1995}}

@misc{NIST_XPS,
  author    = {Naumkin, A. V. and Kraut-Vass, A. and Gaarenstroom, S. W. and Powell, C. J.},
  title     = {{NIST X-ray Photoelectron Spectroscopy Database, NIST Standard Reference Database 20, Version 4.1}},
  publisher = {National Institute of Standards and Technology},
  address   = {Gaithersburg, MD},
  year      = {2012},
  doi       = {10.18434/T4T88K}
}

\clearpage
\onecolumngrid

\setcounter{figure}{0}
\renewcommand{\thefigure}{S\arabic{figure}}

\renewcommand{\theHfigure}{S\arabic{figure}}

\section*{Supplemental Material}

This Supplemental Material provides supporting characterization and quantitative analysis for the substrate-dependent thermal evolution of 100-nm-thick Pt$_{0.9}$Ni$_{0.1}$ films with a 10-nm Zr adhesion layer on sapphire and langasite during sequential high-vacuum annealing from room temperature to 800\,\degC{}. Figure~S1 presents the as-deposited surface morphology and X-ray photoelectron spectroscopy characterization. Figure~S2 documents the ImageJ segmentation procedure used to extract projected feature area and shape parameters from the AFM topographs. Figure~S3 examines the approximate self-similarity of the normalized feature-area distributions in the low-temperature granular regime. Figures~S4 and S5 provide the full temperature-dependent distributions of feature area, circularity, and aspect ratio for PtNi/Langasite and PtNi/Sapphire, respectively. \begin{figure}[!htb] \centering \includegraphics[width=1.0\textwidth]{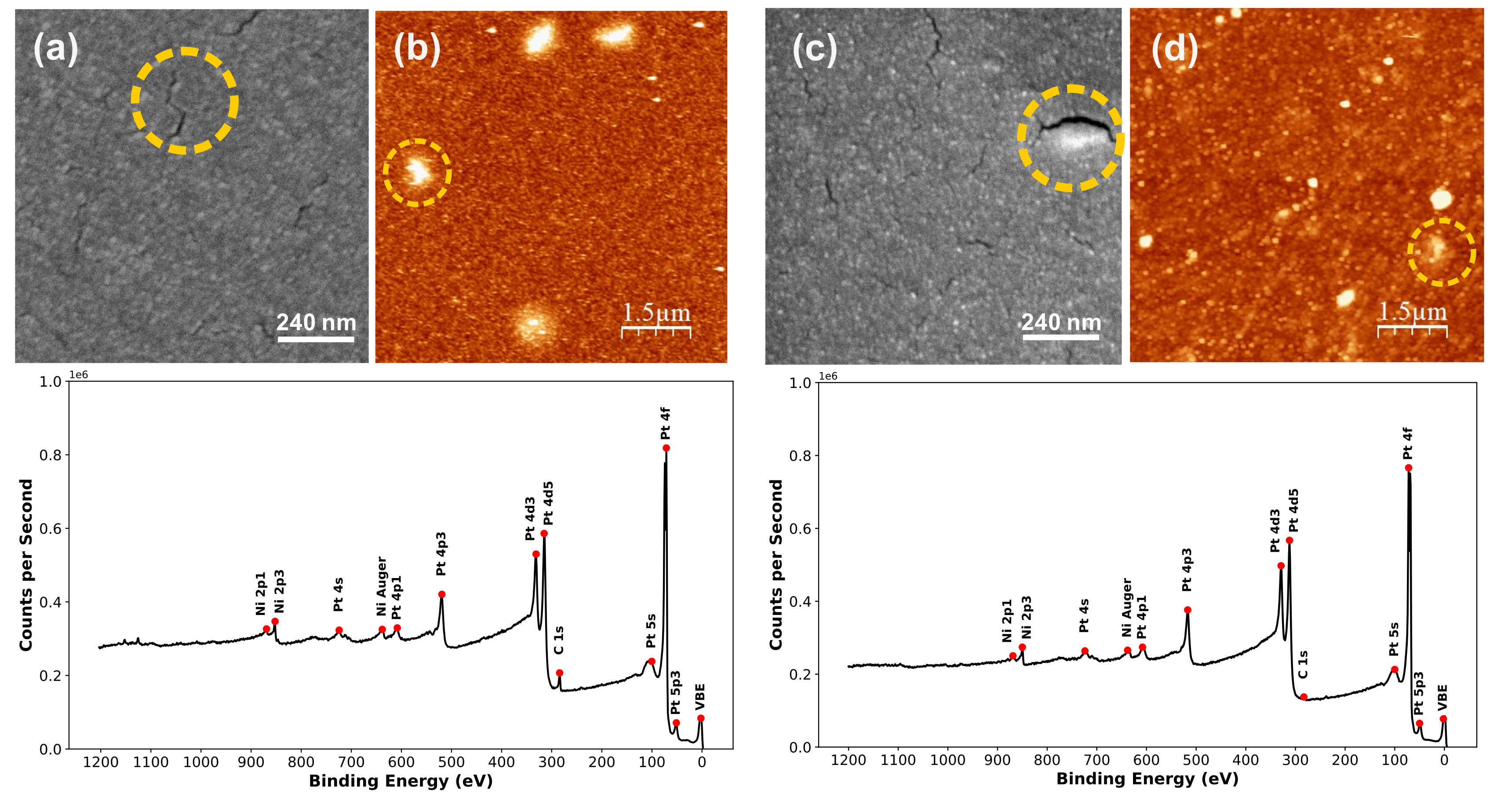} \caption{As-deposited surface characterization of PtNi/Zr films on langasite (left column) and sapphire (right column). Top panels: AFM topographs acquired at room temperature showing the continuous granular morphology on both substrates. Both films show continuous granular coverage with a small number of crack-like surface features. Bottom panels: X-ray photoelectron spectroscopy survey spectra acquired prior to annealing. Both spectra show Pt~4$f$ (71.0 and 74.3~eV), Ni~2$p$ ($\sim$852~eV), Pt~4$d$, C~1$s$ ($\sim$285~eV, adventitious), and valence-band features. The Pt~4$f$ doublet and Ni~2$p$ peaks appear at binding energies consistent with metallic bonding states, confirming the intended alloy composition. The weak C~1$s$ signal is attributed to surface contamination from atmospheric exposure during sample transfer.} \label{fig:xps} \end{figure} \begin{figure}[!htbp] \centering \includegraphics[width=0.95\textwidth]{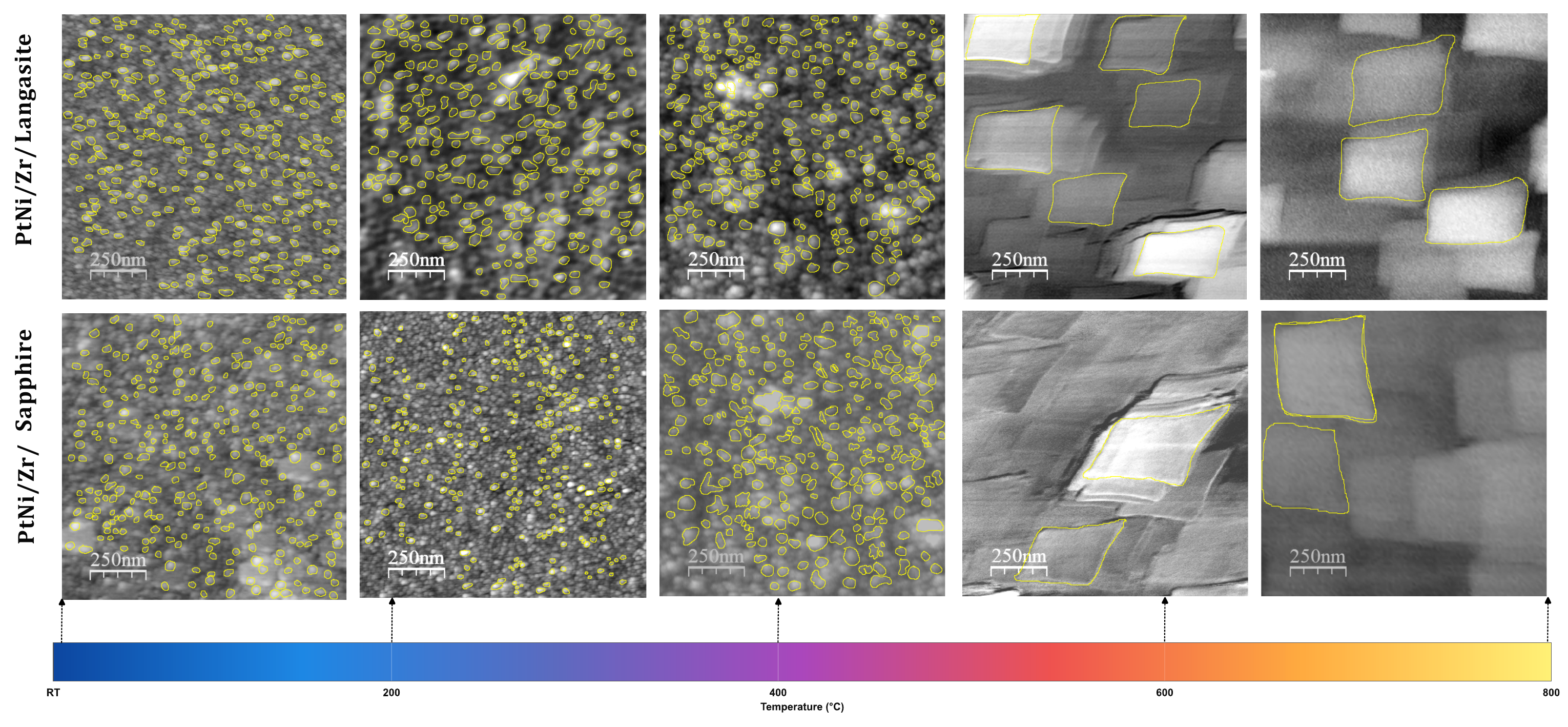} \caption{Segmentation overlays for the ImageJ-based particle analysis of PtNi/Zr films on langasite (top row) and sapphire (bottom row) from RT to 800\degC. Yellow contours mark the features from which area, circularity, Feret diameters, and aspect ratio were extracted. At RT--400\degC, where the morphology is densely granular, 200--700 features were outlined per $1250 \times 1250$~nm$^2$ image using the freehand selection tool in ImageJ. At 600 and 800\degC, where the surface consists of sparse large crystallites, all resolvable features were segmented from multiple scan areas with fields of view ranging from 1250~nm to 5.0~$\mu$m. The same contouring criteria were applied within each substrate and temperature condition.} \label{fig:segmentation} \end{figure}

 \begin{figure}[!htbp] \centering \includegraphics[width=0.95\columnwidth]{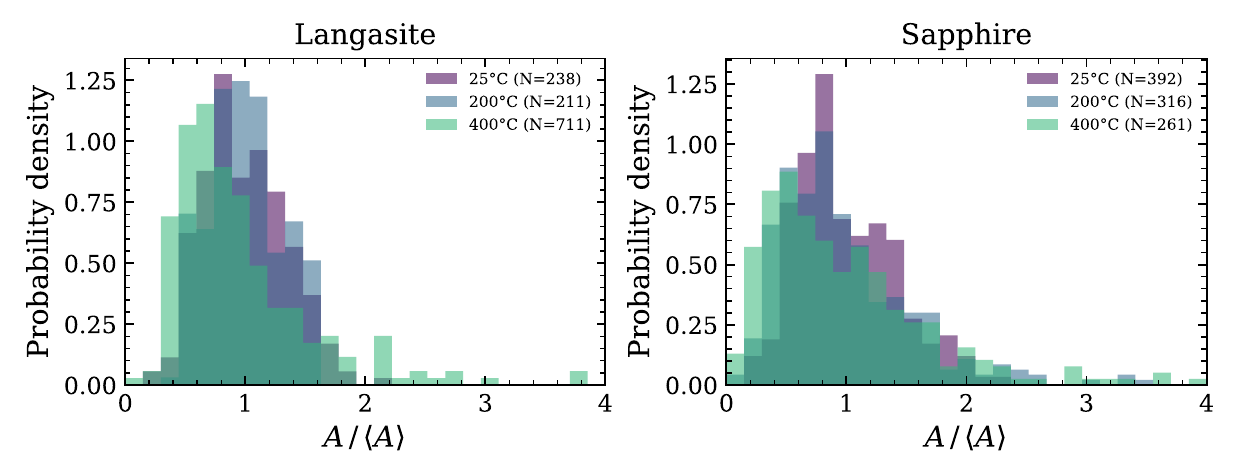} \caption{Normalized area distributions $A/\langle A\rangle$ at RT, 200\degC, and 400\degC{} for PtNi/Langasite (left) and PtNi/Sapphire (right). At these temperatures, where the morphology remains densely granular and the segmented feature populations are large ($N>200$), the rescaled distributions are broadly comparable in shape, consistent with approximate dynamic scaling during granular-state coarsening. The 600 and 800\degC{} states are excluded from this comparison because their substantially smaller feature populations and qualitatively different large-feature morphology make direct assessment of low-temperature self-similarity inappropriate.} \label{fig:selfsimilar} \end{figure}

 \clearpage

 \begin{figure}[!htbp] \centering \includegraphics[width=0.6\textwidth]{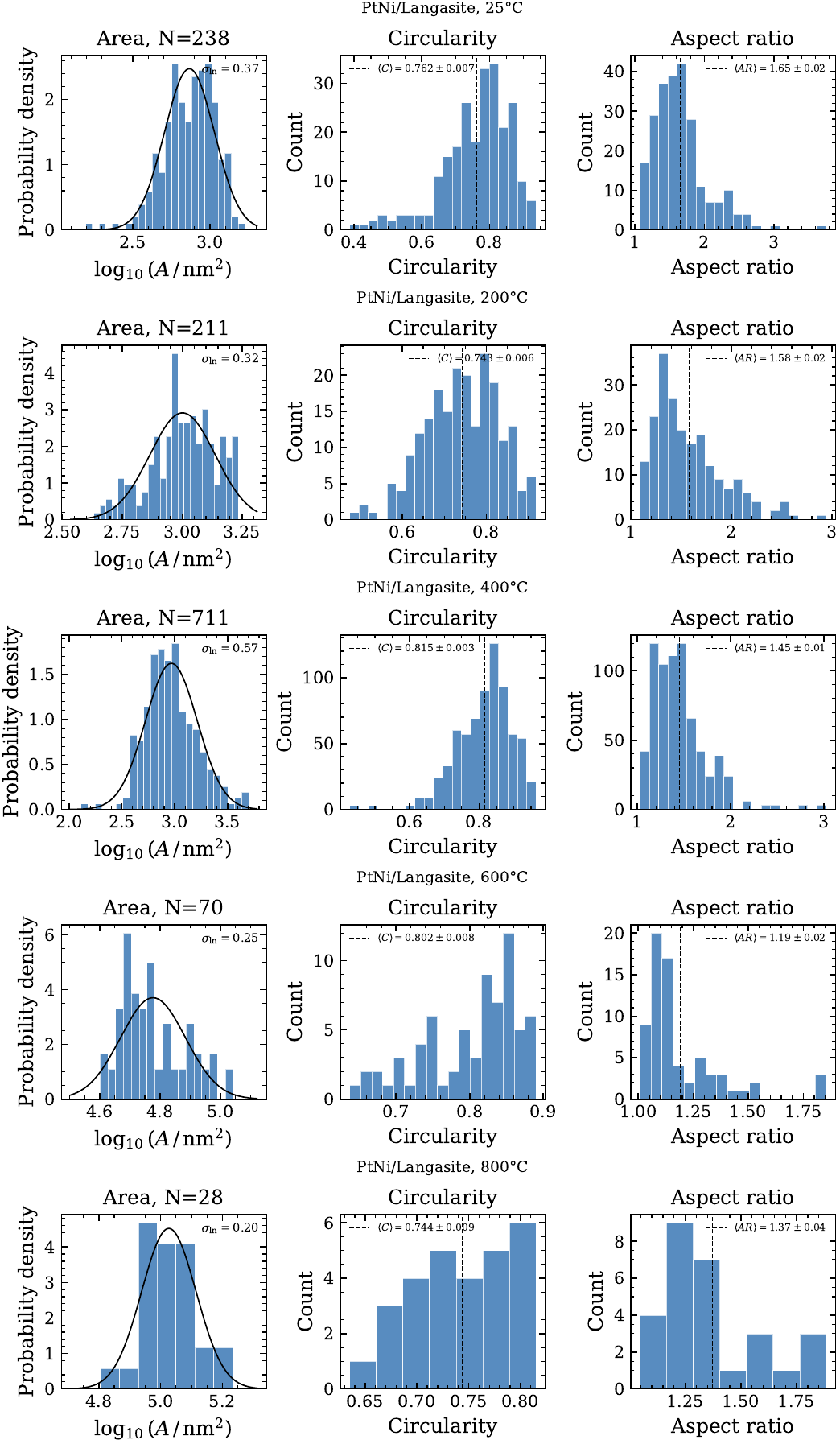} \caption{Per-temperature shape analysis for PtNi/Langasite across the annealing sequence from RT to 800\degC. Each row corresponds to one annealing temperature. Left column: probability-density histograms of $\log_{10}(A/\mathrm{nm}^2)$ with log-normal fits (solid curves) and fitted log-normal width $\sigma_{\ln}$. The log-normal width decreases from $\sigma_{\ln} = 0.37$ at RT to 0.25 at 600\degC, reflecting the narrowing of the area distribution as the large-feature population replaces the broad granular ensemble. Center column: circularity distributions with the mean $\pm$ standard error of the mean indicated by dashed lines. Circularity increases from 0.76 at RT to 0.82 at 400\degC{} and then decreases at 800\degC, consistent with a non-monotonic shape evolution during coarsening and faceting. Right column: aspect ratio distributions showing that the features become more compact (AR closer to unity) at 600\degC{} and then slightly more elongated again at 800\degC. The number of segmented features $N$ is indicated for each temperature.} \label{fig:langasite_detail} \end{figure} 

 \begin{figure}[!htbp] \centering \includegraphics[width=0.6\textwidth]{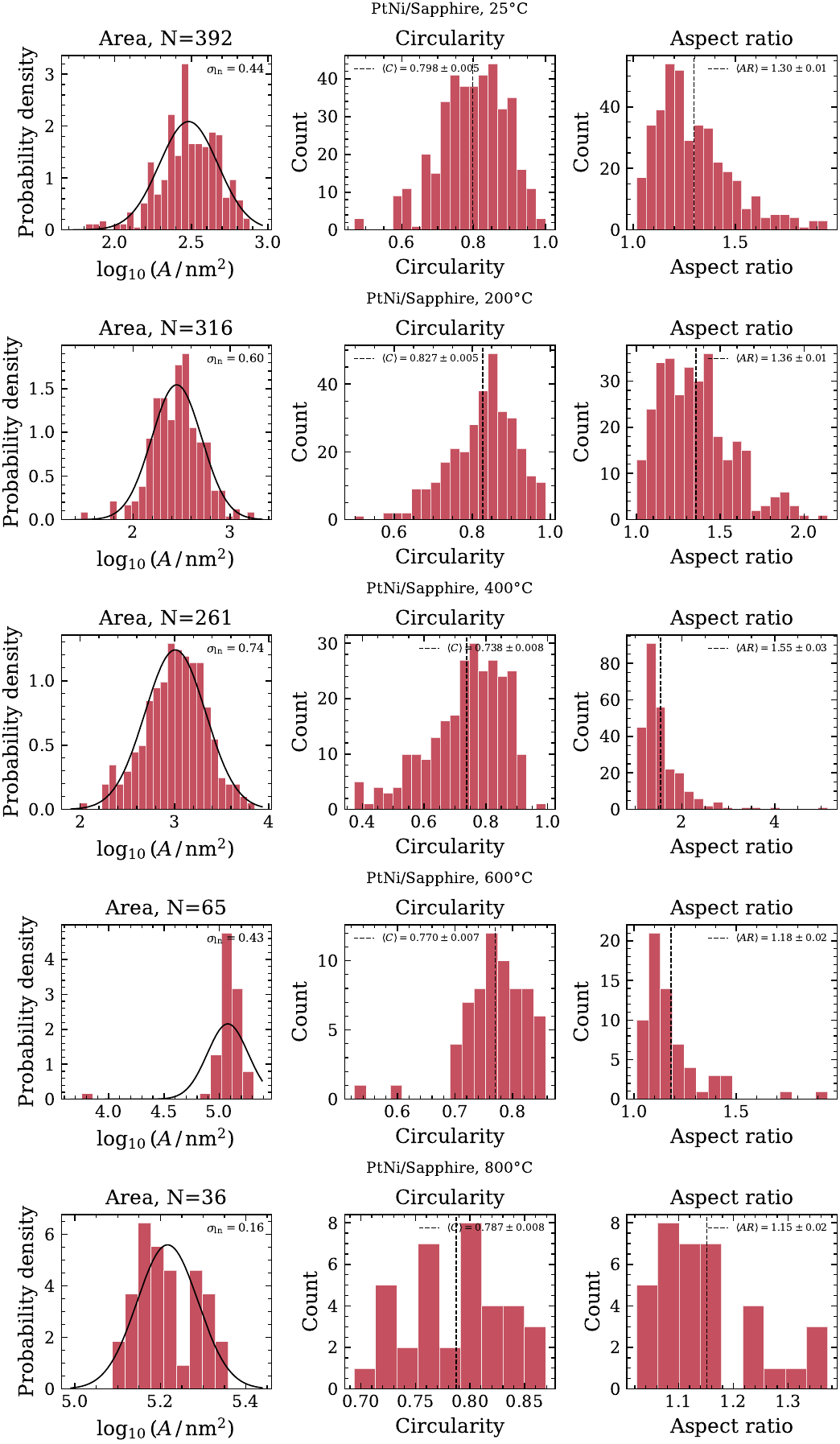} \caption{Per-temperature shape analysis for PtNi/Sapphire. The trends are qualitatively similar to langasite but with quantitative differences. The initial area distribution is narrower ($\sigma_{\ln} = 0.44$) and centered at smaller feature areas ($\langle A\rangle = 332$~nm$^2$). The 400\degC{} distribution is notably broader ($\sigma_{\ln} = 0.74$), reflecting more heterogeneous coarsening. At 600\degC, sapphire develops the larger mean area $(126.0 \pm 3.7)\times10^{3}$~nm$^2$ versus $(61.7 \pm 2.0)\times10^{3}$~nm$^2$ on langasite and shows a moderate reduction in aspect ratio, consistent with the formation of large compact features. By 800\degC, the aspect ratio remains close to its 600\degC{} value (AR $= 1.15$), indicating that the late-stage features shapes have largely stabilized on sapphire.} \label{fig:sapphire_detail} \end{figure}

\end{document}